\documentclass[12pt]{article}

\usepackage{amsmath,amsfonts,amssymb}

\usepackage{times}
\usepackage{bm}
\usepackage{natbib}

\usepackage[plain,noend]{algorithm2e}

\usepackage{xcolor}
\usepackage{graphicx}

\usepackage[margin=1in]{geometry}

\newcommand{\argmin}{\operatorname*{argmin}}
\newcommand{\RR}{\mathbb{R}}
\newtheorem{theorem}{Theorem}

\newtheorem{lemma}{Lemma}

\newtheorem{assumption}{Assumption}

\newtheorem{corollary}{Corollary}

\newtheorem{example}{Example}

\begin{document}

\title{A Double Penalty Model for Interpretability}

\author{Wenjia Wang\\The Statistical and Applied Mathematical Sciences Institute,
Durham, NC \\wenjia.wang234@duke.edu\\ \and Yi-Hui Zhou\\North Carolina State University, Raleigh, NC\\ yihui$\_$zhou@ncsu.edu}

\maketitle

\begin{abstract}
Modern statistical learning techniques have often emphasized prediction performance over interpretability, giving rise to ``black box" models that may be difficult to understand, and to generalize to other settings. We conceptually divide a prediction model into interpretable and non-interpretable portions, as a means to produce models that are highly interpretable with little loss in performance. Implementation of the model is achieved by considering separability of the interpretable and non-interpretable portions, along with a doubly penalized procedure for model fitting. We specify conditions under which convergence of model estimation can be achieved via cyclic coordinate ascent, and the consistency of model estimation holds. We apply the methods to datasets for microbiome host trait prediction and a diabetes trait, and discuss practical tradeoff diagnostics to select models with high interpretability.
\end{abstract}

\textit{keywords}:
Double penalty model; Interpretability; Partially linear model; Separability.

\section{Introduction}

Much of machine learning development has focused on prediction accuracy as a primary criterion \citep{abbott2014applied}. However, recent commentary has emphasized interpretability of models, both to understand underlying relationships and to improve generalizability \citep{doshi2017towards}. Part of the difficulty in moving forward with an emphasis on interpretability is the lack of guiding theory. In addition, the concept of ``interpretability" can be subjective. For example, sparse regression models may be considered intrepretable because there are few coefficients to consider, but non-sparse models (e.g. ridge regression) may also be simple to express and manipulate.

In this paper, we introduce a number of concepts to formalize inherent tradeoffs in model interpretability. A prediction rule is divided into interpretable and uninterpretable portions/functions, as defined by the investigator. The definition and distinction between the functions is arbitrary, and left to the investigator. The key theoretical questions mainly concern consistency and identifiability, which are partly determined by the concept of function separability.  Practical considerations include the development of iterative fitting algorithms for the interpretable/uninterpretable portions, which may be valuable even when separability cannot be established. We emphasize that, in our framework, high interpretablity may come at the cost of prediction accuracy, but a modest loss in accuracy may be worth the gain in interpretability. We illustrate the methods using numerical examples and application to a human microbiome prediction problem.

\section{A double penalty model and a fitting algorithm}\label{secdpm}

Consider a function of interest $h$, which can be expressed by
\begin{align*}
   h(x) = f^*(x) + g^*(x),\quad \forall x\in \Omega,
\end{align*}
where $f^* \in \mathcal{F}$ and $g^* \in \mathcal{G}$ are two unknown functions with known function classes $\mathcal{F}$ and $\mathcal{G}$, respectively. Following the motivation for this work, we suppose that the function class $\mathcal{F}$ consists of functions that are ``easy to interpret," for example, linear functions. We further suppose that $\mathcal{G}$ is judged to be uninterpretable, for example, the output from a random forest procedure. Suppose we observe data $(x_i,y_i)$, $i=1,...,n$ with $y_i = h(x_i) + \epsilon_i$, where $x_i\in \Omega$ and $\epsilon_i$'s are i.i.d. random error with mean zero and finite variance. The goal of this work is to specify or estimate $f^*$ and $g^*$.

Obviously, it is not necessary that $f^*$ and $g^*$ are unique, or can be statistically identified. Regardless of the identifiablity of $f^*$ and $g^*$,  we propose the following double penalty model for fitting, \begin{align}\label{ourPMmodel}
    (\hat f, \hat g) = \argmin_{f\in \mathcal{F}, g\in \mathcal{G}}\frac{1}{n}\sum_{i=1}^n(y_i - f(x_i) - g(x_i))^2 + L_f(f) + L_g(g),
\end{align}
where $L_f$ and $L_g$ are convex penalty functions on $f$ and $g$, and $\hat f$ and $\hat g$ are estimators of $f^*$ and $g^*$, respectively. Under some circumstances, if $f^*$ and $g^*$ can be statistically identified, by using appropriate penalty functions $L_f$ and $L_g$, we can obtain consistent estimators $\hat f$ and $\hat g$ of $f^*$ and $g^*$, respectively. Even if $f^*$ and $g^*$ are nonidentifiable, by using the two penalty functions $L_f$ and $L_g$, the relative contributions of the interpretable and non-interpretable to a final prediction rule can be controlled.

Directly solving \eqref{ourPMmodel} may be difficult, because $L_f(f)$ and $L_g(g)$ may be partly confounded. Here we describe an iterative algorithm to solve the optimization problem in \eqref{ourPMmodel}.

\begin{algorithm}\label{alg:itedou}
\caption{Iterative algorithm}
\hspace*{24pt}
    \textbf{Input:} Data $(x_i,y_i)$, $i=1,...,n$, function classes $\mathcal{F}$ and $\mathcal{G}$, and functions $L_f$ and $L_g$.\\
   \hspace*{24pt} Set $m = 1$.  Let $f_0 = \argmin_{f\in \mathcal{F}}1/n\sum_{i=1}^n(y_i - f (x_i))^2 + L_f(f).$\\
  \hspace*{24pt} \textbf{While} Stopping criteria are not satisfied \textbf{do} \\\
   \hspace*{42pt} Solve
   \begin{align}
    g_m & = \argmin_{g\in \mathcal{G}}\frac{1}{n}\sum_{i=1}^n(y_i - f_{m-1} (x_i) - g(x_i))^2 + L_g(g),\text{ and }\label{gmstep}\\
        f_{m} & = \argmin_{f\in \mathcal{F}}\frac{1}{n}\sum_{i=1}^n(y_i - f (x_i) - g_{m}(x_i))^2 + L_f(f).\label{fmstep}
\end{align}
   \qquad \qquad Set $m = m +1$.\\\
\hspace*{24pt}  \textbf{return} $f_{m}$ and $g_m$.
\end{algorithm}

In Algorithm \ref{alg:itedou}, for each iteration, two separated optimization problems \eqref{fmstep} and \eqref{gmstep} are solved with respect to $f$ and $g$, respectively. The idea of Algorithm \ref{alg:itedou} is similar to the coordinate descent method, which minimizes the objective function with respect to each coordinate direction at a time. The minimization in equations \eqref{gmstep} and \eqref{fmstep} ensures that the function
\begin{align*}
    \frac{1}{n}\sum_{i=1}^n(y_i - f_m(x_i) - g_m(x_i))^2 + L_f(f_m) + L_g(g_m)
\end{align*}
decreases as $m$ increases. One can stop Algorithm \ref{alg:itedou} after reaching a fixed number of iterations or no further improvement of function values can be made.

\section{Separable function classes}\label{secsepcls}
Suppose $f^*$ and $g^*$ can be statistically specified. It can be seen that $\mathcal{F}\cap \mathcal{G} \subset \{0\}$, for otherwise $f^* + w\in \mathcal{F}$ and $g^* - w\in \mathcal{G}$ would be another decomposition of $h$ for any $w\in \mathcal{F}\cap \mathcal{G}$. Furthermore, we would expect that $\mathcal{F}$ and $\mathcal{G}$ are non-overlapping, in the sense defined here. We define $\mathcal{F}$ and $\mathcal{G}$ as \textit{$L_2$-separable} if there exists $\theta_1\in [0,1)$ such that for any functions $f\in \mathcal{F}$ and $g \in \mathcal{G}$,
\begin{align}\label{ineqsepecond}
  |\langle f,g \rangle_2|  \leq \theta_1\|f\|_{L_2}\|g\|_{L_2},
\end{align}
where $\|f\|_{L_2}$ denotes the $L_2$ norm of a function $f\in L_2(\Omega)$, and $\langle f,g \rangle_2$ denotes the inner product of functions $f,g\in L_2(\Omega)$. In fact, if two function classes are $L_2$-separable, then $f^*$ and $g^*$ are unique, as stated in the following lemma.

\begin{lemma}\label{lemfstarunique}
Suppose \eqref{ineqsepecond} is true for any functions $f\in \mathcal{F}$ and $g \in \mathcal{G}$, then $f^*$ and $g^*$ are unique, up to a difference on a measure zero set.
\end{lemma}
Lemma \ref{lemfstarunique} implies that if two function classes are $L_2$-separable, then $f^*$ and $g^*$ are identifiable.

\medskip
We start with convergence analysis of Algorithm \ref{alg:itedou} applying to separable function classes, and then provide some examples of separable function classes with theoretical properties.

\subsection{Convergence of Algorithm \ref{alg:itedou}}

The convergence results depend on the separability with respect to the empirical norm. Define the empirical inner product as
$$\langle f, g \rangle_n = \frac{1}{n}\sum_{i=1}^n f(x_i)g(x_i)$$ for two functions $f$ and $g$, and $\|g\|_n^2 = \langle g, g \rangle_n$ as the empirical norm of function $g$. Analogous to the earlier definition, we say $\mathcal{F}$ and $\mathcal{G}$ are separable with respect to the empirical norm, if there exists $\theta_2\in [0,1)$ such that for any functions $f\in \mathcal{F}$ and $g \in \mathcal{G}$,
\begin{align}\label{ineqsepecondnnorm}
  |\langle f,g \rangle_n|  \leq \theta_2\|f\|_n\|g\|_n.
\end{align}
Now we are ready to present the convergence results, under the condition that $\mathcal{F}$ and $\mathcal{G}$ are separable with respect to the empirical norm.
\begin{theorem}\label{cons11}
Suppose \eqref{ineqsepecondnnorm} is true for any functions $f\in \mathcal{F}$ and $g \in \mathcal{G}$. Let $f_m$ and $g_m$ be defined as in \eqref{fmstep} and \eqref{gmstep}, respectively. We have
\begin{align*}
    \|f_m - \hat f\|_n + \|g_m - \hat g\|_n \leq \theta_2^{2m-6}(\|f_1 - \hat f\|_n + \|g_1 - \hat g\|_n),
\end{align*}
and $(L_f(f_m),L_g(g_m)) \rightarrow (L_f(\hat f),L_g(\hat g))$, as $m$ goes to infinity. Here we use the convention $1/0 = \infty$.
\end{theorem}
It can be seen that in Theorem \ref{cons11}, if $\mathcal{F}$ and $\mathcal{G}$ are separable with respect to the empirical norm, Algorithm \ref{alg:itedou} achieves a linear convergence. The parameter $\theta_2$ determines the convergence speed.

Since the empirical norm is close to $L_2$ norm as the sample size increases \citep{geer2000empirical}, it can be expected that if $\mathcal{F}$ and $\mathcal{G}$ are $L_2$-separable, then $\mathcal{F}$ and $\mathcal{G}$ are separable with respect to the empirical norm. Furthermore, it can be shown that $\theta_2$ is close to $\theta_1$ in \eqref{ineqsepecond}. These claims are verified in the following lemma.

\begin{lemma}\label{sepeequi}
Let $\mathcal{F}$ and $\mathcal{G}$ be $L_2$-separable satisfying \eqref{ineqsepecond}. Suppose $x_i$'s are uniformly distributed on $\Omega$. For $t\in [C_1, C_2n]$, $n\geqslant N$, any $\alpha \in (0,1/4)$, and any functions $f\in \mathcal{F}$ and $g\in\mathcal{G}$ satisfying $\|f\|_n \|g\|_n > C_3t^{1/4}n^{-1/4 + \alpha}$, with probability at least $1 - C_4\exp(-t)$, $|\langle f,g\rangle_n| \leq (\theta_1 + C_5n^{-\alpha}) \|f\|_n\|g\|_n$, where $N$ and $C_i$'s are constants only depending on $\mathcal{F}$, $\mathcal{G}$ and $\Omega$.
\end{lemma}

By Theorem \ref{cons11} and Lemma \ref{sepeequi}, we have the following corollary.

\begin{corollary}\label{coroconalg}
Suppose $\mathcal{F}$ and $\mathcal{G}$ are $L_2$-separable satisfying \eqref{ineqsepecond}, and $x_i$'s are uniformly distributed on $\Omega$. Let $f_m$ and $g_m$ be as in Theorem \ref{cons11}. For $t\in [C_1, C_2n]$, $n\geqslant N$, and any $\alpha \in (0,1/4)$, with probability at least $1 - C_3\exp(-t)$,  we have either $\|f_m - \hat f\|_n^2 + \|g_m - \hat g\|_n^2 \leq C_4t^{1/4}n^{-1/4 + \alpha}$, or
\begin{align*}
    \|f_m - \hat f\|_n + \|g_m - \hat g\|_n \leq \bigg(\theta_1 + C_5n^{-\alpha}\bigg)^{2m-6}(\|f_1 - \hat f\|_n + \|g_1 - \hat g\|_n),
\end{align*}
and $(L_f(f_m),L_g(g_m)) \rightarrow (L_f(\hat f),L_g(\hat g))$,
as $m$ goes to infinity, where $N$ and $C_i$'s are constants only depending on $\mathcal{F}$, $\mathcal{G}$ and $\Omega$.
\end{corollary}
Corollary \ref{coroconalg} shows that if $\theta_1$ in \eqref{ineqsepecond} is small and sample size is relatively large, then Algorithm \ref{alg:itedou} converges fast. In particular, if for any functions $f\in\mathcal{F}$ and $g\in \mathcal{G}$, $\langle f,g\rangle_2 = 0$, a few iterations of Algorithm \ref{alg:itedou} are sufficient to get a good numerical solution to \eqref{ourPMmodel}.

\subsection{Finite dimensional function classes}\label{subsecsubmodel1}
We start with the easiest case. Suppose two function classes $\mathcal{F}$ and $\mathcal{G}$ have finite dimensions. To be specific, suppose
\begin{align*}
    \mathcal{F} & = \bigg\{f = \sum_{k=1}^{d_1} \alpha_k \phi_k: \alpha=(\alpha_1,...,\alpha_{d_1})^T\in \RR^{d_1}, \|f\|_{L_2} \leq R_f \bigg\}, \text{ and }\\
    \mathcal{G} & = \bigg\{g = \sum_{j=1}^{d_2} \beta_j \varphi_j: \beta=(\beta_1,...,\beta_{d_2})^T\in \RR^{d_2}, \|g\|_{L_2} \leq R_g \bigg\},
\end{align*}
where $\phi_k,\varphi_j$'s are known functions defined on a compact set $\Omega$, and $R_f$ and $R_g$ are known constants. Furthermore, assume $\mathcal{F}$ and $\mathcal{G}$ are $L_2$-separable. Since the dimension of each function class is finite, we can use the least squares method to estimate $f^*$ and $g^*$, i.e.,
\begin{align}\label{oursubmodel1}
    (\hat f, \hat g) = \argmin_{f\in \mathcal{F}, g\in \mathcal{G}}\frac{1}{n}\sum_{i=1}^n(y_i - f(x_i) - g(x_i))^2.
\end{align}
By applying standard arguments in the theory of Vapnik-Chervonenkis subgraph class \citep{geer2000empirical}, the consistency of $\hat f$ and $\hat g$ holds. We do not present detailed discussion for the conciseness of this paper.

Although the exact solution to the optimization problem in \eqref{oursubmodel1} is available, we can still use Algorithm \ref{alg:itedou} to solve it. By comparing the exact solution with numeric solution obtained by Algorithm \ref{alg:itedou}, we can study the convergence rate of Algorithm \ref{alg:itedou} via numerical simulations. The detailed numerical studies of the convergence rate is provided in Section \ref{secconvegnum}.



\subsection{A generalization of partially linear models}\label{subsecsubmodel2}
In this subsection, we consider a generalization of partially linear models, where the responses can be expressed as
\begin{align}\label{PLMmodel}
    y = x^T\beta + g(t) + \epsilon.
\end{align}
In the partially linear models \eqref{PLMmodel}, $\beta\in \RR^p$ is a vector of regression coefficients associated with $x$, $g$ is an unknown function of $t$ with some known smoothness, which is usually a one dimensional scalar, and $\epsilon$ is a random noise. The partially linear model \eqref{PLMmodel} can be estimated by the partial spline estimator \citep{wahba1984partial,heckman1986spline}, partial residual estimator \citep{speckman1988kernel,chen1988convergence}, or SCAD-penalized regression \citep{xie2009scad}.

In this work, we consider a more general model. Suppose we observe data $y_i$ on $x_i\in \Omega = [0,1]^p$ for $i=1,...,n$, where
\begin{align}\label{oursubmodel2}
    y_i = x_i^T\beta^* + g^*(x_i) + \epsilon_i,
\end{align}
and $\epsilon_i$'s are i.i.d. random errors with mean zero and finite variance. We assume that the function $g^* \in H^{\nu}(\Omega)$, where $H^{\nu}(\Omega)$ is the Sobolev space with known smoothness $\nu$. This is a standard assumption in nonparametric regression, see \cite{gu2013smoothing,geer2000empirical} for examples. It is natural to define the two function classes by
\begin{align*}
    \mathcal{F} = \left\{f(x) = x^T\beta: \beta=(\beta_1,...,\beta_{p})^T\in \RR^{p}, \|\beta\|_2\leq R, x\in \Omega \right\}
\end{align*}
and $\mathcal{G} = H^{\nu}(\Omega)$,
where $\|\cdot\|_2$ denotes the Euclidean distance, and $R$ is a known constant. In practice, we can choose a sufficient large $R$ such that $\|\beta\|_2\leq R$ is fulfilled. Note that in our generalization of the partially linear model, we \textit{cannot} separate variables $x$ and $t$, which is not the case of the partially linear model \eqref{PLMmodel}. It can be seen that $\beta^*$ and $g^*$ are non-identifiable because $\mathcal{F} \subset \mathcal{G}$. Furthermore, $\mathcal{F}$ is more interpretable compared with $\mathcal{G}$ because it is linear.

In order to uniquely identify $\beta^*$ and $g^*$, we need to restrict function class $\mathcal{G}$ such that $\mathcal{F}$ and $\mathcal{G}$ are separable. This can be done by applying a newly developed approach, employing the \textit{projected kernel} \citep{tuo2019adjustments}. Let $e_k$, $k=1,...,p$ be an orthonormal basis of $\mathcal{F}$. Then $\mathcal{F}$ can be defined as a linear span of the basis $\{e_1,...,e_p\}$, and the projection of a function $w\in\mathcal{G}$ on $\mathcal{F}$ is given by
\begin{align}\label{L2project}
    \mathcal{P}_\mathcal{F}w = \sum_{k=1}^p\langle w,e_k\rangle_2 e_k.
\end{align}
The perpendicular component is
\begin{align}\label{L2projectperp}
    \mathcal{P}_\mathcal{F}^\perp w = w - \mathcal{P}_\mathcal{F}w.
\end{align}
By \eqref{L2project} and \eqref{L2projectperp}, we can split $\mathcal{G}$ into two perpendicular classes as $\mathcal{F}$ and $\mathcal{F}^\perp$, where $\mathcal{F}^\perp = \{w_1 = \mathcal{P}_\mathcal{F}^\perp w, w\in \mathcal{G}\}$.
Let $h = x^T\beta^* + g^*(x)$, where $g^* \in \mathcal{F}^\perp$. Since $\mathcal{F}$ and $\mathcal{F}^\perp$ are perpendicular, they are $L_2$-separable. By Lemma \ref{lemfstarunique}, $\beta^*$ and $g^*$ are unique. However, in practice it is usually difficult to find a function $g^* \in \mathcal{F}^\perp$ directly. We propose using projected kernel ridge regression, which depends on the reproducing kernel Hilbert space generated by the projected kernel.

Let $\Psi$ be the (isotropic) Mat\'ern family \citep{stein1999interpolation}, defined by
\begin{align}\label{kernelfunctionPsi}
	\Psi(s,t)=\frac{(2\sqrt{\nu- p/2}\phi \| s-t\|)^{\nu- p/2}}{\Gamma(\nu- p/2)2^{\nu- p/2-1}} K_{\nu- p/2}(2\sqrt{\nu- p/2}\phi\| s-t\|),
\end{align}
where $K_{\nu - p/2}$ is the modified Bessel function of the second kind, and $\phi$ is the range parameter. We use $\mathcal{N}_{\Psi}(\Omega)$ to denote the \textit{reproducing kernel Hilbert space} generated by $\Psi$, and $\|\cdot\|_{\mathcal{N}_{\Psi}(\Omega)}$ to denote the norm of $\mathcal{N}_{\Psi}(\Omega)$. By Corollary 10.48 in \cite{wendland2004scattered}, $\mathcal{G}$ coincides with $\mathcal{N}_{\Psi}(\Omega)$. The reproducing kernel Hilbert space generated by the projected kernel can be defined in the following way. Define the linear operators $\mathcal{P}_\mathcal{F}^{(1)}$ and $\mathcal{P}_\mathcal{F}^{(2)}$: $L_2(\Omega \times \Omega)\rightarrow L_2(\Omega \times \Omega)$ as
\begin{align*}
    \mathcal{P}_\mathcal{F}^{(1)}(u)(x,y) = \sum_{k=1}^p e_k(x)\int_\Omega u(s,y)e_k(s)ds,\\
    \mathcal{P}_\mathcal{F}^{(2)}(u)(x,y) = \sum_{k=1}^p e_k(y)\int_\Omega u(x,t)e_k(t)dt,
\end{align*}
for $u\in L_2(\Omega \times \Omega)$. The projected kernel of $\Psi$ can be defined by
\begin{align}\label{projectkdef}
    \Psi_{\mathcal{F}} = \Psi - \mathcal{P}_\mathcal{F}^{(1)}\Psi - \mathcal{P}_\mathcal{F}^{(2)}\Psi + \mathcal{P}_\mathcal{F}^{(1)}\mathcal{P}_\mathcal{F}^{(2)}\Psi.
\end{align}
The function class $\mathcal{F}^\perp$ then is equivalent to the reproducing kernel Hilbert space generated by $\Psi_{\mathcal{F}}$, denoted by $\mathcal{N}_{\Psi_{\mathcal{F}}}(\Omega)$, and the norm is denoted by $\|\cdot\|_{\mathcal{N}_{\Psi_{\mathcal{F}}}(\Omega)}$. For detailed discussion and properties of $\Psi_{\mathcal{F}}$ and $\mathcal{N}_{\Psi_{\mathcal{F}}}(\Omega)$, we refer to \cite{tuo2019adjustments}.

By using the projected kernel of $\Psi$, the double penalty model is
\begin{align}\label{oursubmodel2p}
    (\hat \beta, \hat g) = \argmin_{\beta \in \RR^p, g\in \mathcal{N}_{\Psi_{\mathcal{F}}}(\Omega)} \|y - x^T\beta - g\|_n^2 + \lambda \|g\|_{\mathcal{N}_{\Psi_{\mathcal{F}}}(\Omega)}^2,
\end{align}
where $(\hat \beta, \hat g)$ are estimators of $(\beta^*, g^*)$. In practice, we can use generalized cross validation (GCV) to choose the tuning parameter $\lambda$ \citep{tuo2019adjustments,wahba1990spline}. If the tuning parameter $\lambda$ is chosen properly, we can show that $(\hat \beta, \hat g)$ are consistent, as stated in the following theorem. In the rest of this paper, we use the following notation. For two positive sequences $a_n$ and $b_n$, we write $a_n\asymp b_n$ if, for some constants $C,C'>0$, $C\leq a_n/b_n \leq C'$.
\begin{theorem}\label{submodel2thmconsist}
Suppose $x_i$'s are uniformly distributed on $\Omega$, and the noise $\epsilon_i$'s are i.i.d. sub-Gaussian, i.e., satisfying $K^2\mathbb{E}\exp(|\epsilon_i|^2/K^2) -1\leq \sigma_0^2$ for some constants $K$ and $\sigma_0^2$, and all $i=1,...,n$. Futhermore, assume Assumption \ref{assumentrf} in Appendix holds. If $\lambda \asymp n^{-2\nu/(2\nu + p)}$, we have
\begin{align*}
    \|\hat g - g^*\|_{L_2}^2 = O_{P}(n^{-\frac{2\nu}{2\nu + p}}), \|\hat \beta - \beta^*\|_2^2 = O_{P}(n^{-\frac{2\nu}{2\nu + p}}).
\end{align*}
\end{theorem}
Theorem \ref{submodel2thmconsist} shows that the double penalty model \eqref{oursubmodel2p} can provide consistent estimators of $\beta^*$ and $g^*$, and the convergence rate of $\|\hat g - g^*\|_{L_2}$ is known to be optimal \citep{stone1982optimal}. The convergence rate of $\|\hat \beta - \beta^*\|_2$ in Theorem \ref{submodel2thmconsist} is slower than the convergence rate $n^{-1/2}$ in the linear model. We conjecture that the convergence rate of $\|\hat \beta - \beta^*\|_2$ in Theorem \ref{submodel2thmconsist} is optimal since it is influenced by the estimation of $g^*$, which may introduce extra error because functions in $\mathcal{N}_{\Psi_{\mathcal{F}}}(\Omega)$ and $\mathcal{F}$ have the same input space.

In order to solve the optimization problem in \eqref{oursubmodel2p}, we apply Algorithm \ref{alg:itedou}. It can be checked that $\mathcal{F}$ and $\mathcal{N}_{\Psi_{\mathcal{F}}}(\Omega)$ are separable with respect to the empirical norm, as presented in the following corollary.
\begin{corollary}\label{subm2sepcoro}
$\mathcal{F}$ and $\mathcal{N}_{\Psi_{\mathcal{F}}}(\Omega)$ are separable with respect to the empirical norm with probability tending to one, as the sample size goes to infinity.
\end{corollary}
Corollary \ref{subm2sepcoro} is a direct result of Lemma \ref{sepeequi}, thus the proof is omitted. By Theorem \ref{cons11} and Corollary \ref{subm2sepcoro}, the convergence of Algorithm \ref{alg:itedou} can be guaranteed. In each iteration of Algorithm \ref{alg:itedou}, $g_m$ and $f_m$ are solved by
\begin{align*}
g_m = &\argmin_{g\in \mathcal{N}_{\Psi_{\mathcal{F}}}(\Omega)} \|y - x^T\beta_{m-1} - g\|_n^2 + \lambda \|g\|_{\mathcal{N}_{\Psi_{\mathcal{F}}}(\Omega)}^2,\text{ and }\\
\beta_m = &\argmin_{\beta \in \RR^p} \|y - x^T\beta - g_m\|_n^2,
\end{align*}
which have explicit forms as
\begin{align*}
    g_m(x) =& r(x)^T(K+n\lambda)^{-1}(Y - X^T\beta_{m-1}),\text{ and }\\
    \beta_m =& (X^TX)^{-1}X^T(Y-g_m(X)),
\end{align*}
where $r(x) = (\Psi_{\mathcal{F}}(x,x_1),...,\Psi_{\mathcal{F}}(x,x_n))^T$, $K = (\Psi_{\mathcal{F}}(x_j, x_k))_{jk}$, $X = (x_1,...,x_n)^T$, $g_m(X) = (g_m(x_1),...,g_m(x_n))^T$, and $Y = (y_1,...,y_n)^T$. Because $\mathcal{F}$ and $\mathcal{N}_{\Psi_{\mathcal{F}}}(\Omega)$ are orthogonal, a few iterations of Algorithm \ref{alg:itedou} are sufficient to obtain a good numeric solution.

\section{Non-separable function classes}
In Section \ref{secsepcls}, we consider the case that $\mathcal{F}$ and $\mathcal{G}$ are $L_2$-separable, which implies $f^*$ and $g^*$ are statistically identifiable. However, in many practical cases, $\mathcal{F}$ and $\mathcal{G}$ are not $L_2$-separable. Such examples include $\mathcal{F}$ as a linear function class and $\mathcal{G}$ as the function space generated by a neural network.
If $\mathcal{F}$ and $\mathcal{G}$ are not $L_2$-separable, then $f^*$ and $g^*$ are not statistically identifiable. To see this, note that there exist two sequences of functions $\{f'_j\} \subset \mathcal{F}$ and $\{g'_j\} \subset \mathcal{G}$ such that $\|f'_j - g'_j\|_{L_2} \rightarrow 0$. This implies that $(f^*,g^*)$ and $(f^* - f'_j, g^* + g'_j)$ are not statistically identifiable, which implies that we cannot consistently estimate $f^*$ and $g^*$.

Although $\mathcal{F}$ and $\mathcal{G}$ can be not $L_2$-separable, we can still use \eqref{ourPMmodel} to specify $f^*$ and $g^*$. We propose choosing $\mathcal{F}$ with  simple structure and to be ``easy to interpret," and choosing $\mathcal{G}$ to be flexible to improve the prediction accuracy. The tradeoff between interpretation and prediction accuracy can be adjusted by applying different penalty functions $L_f$ and $L_g$. If $L_f$ is large, then \eqref{ourPMmodel} forces $f^*$ to be small and $g^*$ to be large, which indicates that the model is more flexible, but is less interpretable. On the other hand, if $L_g$ is large, then the model is more interpretable, but may reduce the power of prediction.

The theoretical analysis of arbitrary function classes $\mathcal{F}$ and $\mathcal{G}$ varies case by case, and depends on the structure of function classes. We do not discuss theoretical results of specifying non-$L_2$-separable function classes, which are beyond the scope of this paper.

\subsection{Convergence of Algorithm \ref{alg:itedou}}

For non-separable function classes, a stronger condition on penalty functions is needed to show the convergence of Algorithm \ref{alg:itedou}. Let $\|\cdot\|$ be a (semi-)norm of a Hilbert space. A function $L$ is said to be \textit{strongly convex} with respect to (semi-)norm $\|\cdot\|$ if there exists a parameter $\gamma>0$ such that for any $x,y$ in the domain and $t\in [0,1]$,
\begin{align*}
    L(tx + (1-t)y) \leq tL(x) + (1-t)L(y) -\frac{1}{2}\gamma t(1-t)\|x-y\|^2.
\end{align*}
As a simple example, $\|\cdot\|^2$ is strongly convex for any norm $\|\cdot\|$. If $L_f$ or $L_g$ is strongly convex, Algorithm \ref{alg:itedou} converges, as stated in the following theorem.
\begin{theorem}\label{cons22}
Suppose $L_f$ or $L_g$ is strongly convex with respect to the empirical norm with parameter $\gamma>0$. We have
\begin{align*}
    \|f_m - \hat f\|_n + \|g_m - \hat g\|_n \leq \left(\frac{2}{2 + \gamma}\right)^{m-1}(\|f_1 - \hat f\|_n + \|g_1 - \hat g\|_n),
\end{align*}
and $(L_f(f_m),L_g(g_m)) \rightarrow (L_f(\hat f),L_g(\hat g))$,
as $m$ goes to infinity.
\end{theorem}
From Theorem \ref{cons22}, it can be seen that Algorithm \ref{alg:itedou} can also achieve a linear convergence if $L_f$ or $L_g$ is strongly convex, even if $f^*$ and $g^*$ are not statistical identifiable. We only require one penalty function to be strongly convex. The convergence rate depends on the parameter $\gamma$, which measures the convexity of a function. If the penalty function is more convex, i.e., $\gamma$ is larger, then the convergence of Algorithm \ref{alg:itedou} is faster. The strong convexity of the penalty function $L_f$ or $L_g$ can be easily fulfilled, because the square of norm of any Hilbert space is strongly convex. For example, the penalty functions in the ridge regression and the neural networks with fixed number of neuron are strongly convex with respect to the empirical norm.

We summarize
Theorem \ref{cons11} and Theorem \ref{cons22} in the following corollary.
\begin{corollary}\label{cons}
Suppose $\mathcal{F}$ and $\mathcal{G}$ are separable with respect to the empirical norm, or one function of $L_f$ and $L_g$ is strongly convex with respect to the empirical norm. We have
\begin{align*}
    \|f_m - \hat f\|_n + \|g_m - \hat g\|_n \rightarrow 0,
\end{align*}
and $(L_f(f_m),L_g(g_m)) \rightarrow (L_f(\hat f),L_g(\hat g))$,
as $m$ goes to infinity.
\end{corollary}

\section{Numerical examples}\label{secnum}


\subsection{Convergence rate of Algorithm \ref{alg:itedou}}\label{secconvegnum}

In this subsection, we report numerical studies on the convergence rate of Algorithm \ref{alg:itedou}, and verify the convergence rate in Theorem \ref{cons11} is sharp. We consider two finite function classes such that the analytic solution of \eqref{oursubmodel1} is available, as stated in Section \ref{subsecsubmodel1}. By comparing the numeric solution and the analytic solution, we can verify the convergence rate is sharp.


We consider two function classes $\mathcal{F} = \{ f|f(x) = \alpha_1 x, \alpha_1 \in [0, 10]\}$ and $\mathcal{G} = \{g|g(x) = \alpha_2 \sin(\theta x), \alpha_2 \in [0, 10]\}$, where $x\in [0,1]$, $\theta$ is a known parameter which controls the degree of separation of two function classes, i.e., the parameter $\theta_1$ in Lemma \ref{lemfstarunique}.
It is easy to verify that for $f\in \mathcal{F}$ and $g\in \mathcal{G}$,
\begin{align*}
    \bigg|\int_0^1 f(x) g(x) dx \bigg|\leq \frac{2\sqrt{3\theta}|\sin(\theta) - \theta\cos(\theta)|}{\theta^2\sqrt{2\theta - \sin(2\theta)}} \|f\|_{L_2([0,1])}\|g\|_{L_2([0,1])}.
\end{align*}
Let $$\psi(\theta) = \frac{2\sqrt{3\theta}|\sin(\theta) - \theta\cos(\theta)|}{\theta^2\sqrt{2\theta - \sin(2\theta)}}.$$ 
Suppose the underlying function $h(x) = \beta_1^* x + \beta_2^*\sin(\theta x)$ with $(\beta_1^*,\beta_2^*) = (1,3)$. Let $(\hat \beta_1,\hat \beta_2)$ be the solution to \eqref{oursubmodel1}, and $(\beta_{1,m},\beta_{2,m})$ be the values obtained at $m$th iteration of Algorithm \ref{alg:itedou}. By Theorem \ref{cons11},
\begin{align}\label{convereg1eq}
    & \|(\beta_{1,m} - \hat \beta_1)x\|_n + \|(\beta_{2,m} - \hat \beta_2)\sin(\theta x)\|_n \nonumber\\
    \leq & \theta_1^{2m-6}(\|(\beta_{1,1} - \hat \beta_1)x\|_n + \|(\beta_{2,1} - \hat \beta_2)\sin(\theta x)\|_n).
\end{align}
By taking logarithms on both sides of \eqref{convereg1eq}, we have
\begin{align}\label{convereg1eq2}
   & \log(\|(\beta_{1,m} - \hat \beta_1)x\|_n + \|(\beta_{2,m} - \hat \beta_2)\sin(\theta x)\|_n)\nonumber\\
   \leq & \log(\theta_1^{2m-6}(\|(\beta_{1,1} - \hat \beta_1)x\|_n + \|(\beta_{2,1} - \hat \beta_2)\sin(\theta x)\|_n))\nonumber\\
    \approx & 2\log(\psi(\theta)) m + \log(\psi(\theta)^{-6}(\|(\beta_{1,1} - \hat \beta_1)x\|_2 + \|(\beta_{2,1} - \hat \beta_2)\sin(\theta x)\|_2)),
\end{align}
where the approximation is because of Corollary  \ref{coroconalg}. If the convergence rate in Theorem \ref{cons11} is sharp, $\log(\|(\beta_{1,m} - \hat \beta_1)x\|_n + \|(\beta_{2,m} - \hat \beta_2)\sin(\theta x)\|_n)$ is an approximate linear function with respect to $m$ and the slope is close to $2\log(\psi(\theta))$.

In our simulation studies, we choose $\theta=2,3,3.5,4$. We choose the noise $\epsilon \sim N(0,0.1)$, where $N(0,0.1)$ is a normal distribution with mean zero and variance $0.1$. The algorithm stops if the left hand side of \eqref{convereg1eq} is less than $10^{-6}$.
We choose 50 uniformly distributed points as training points. We run 100 simulations and take the average of the regression coefficient and the number of iterations needed for each $\theta$. The results are shown in Table \ref{Tab:simuResults}.


\begin{table}[h]
\centering
\begin{tabular}{c|c|c|c|c|c}
\hline
$\theta$ & $\psi(\theta)$ & $2\log(\psi(\theta))$ &  Regression coefficient & Iteration numbers & Absolute difference\\
\hline
2 & 0.978 & -0.045 & -0.050 & 491.55 & 0.006\\
3 & 0.828 &  -0.378 & -0.419 & 59.02 & 0.040 \\
3.5 & 0.615 & -0.973 &-1.121  & 22.34 & 0.148 \\
4 & 0.304 &-2.383 &-2.624 & 10 & 0.241\\
\hline
\end{tabular}
\caption{The simulation results when the sample size is fixed. The last column show the absolute difference between the third column and the fourth column, given by $|2\log(\psi(\theta))-$ Regression coefficient$|$.}
\label{Tab:simuResults}
\end{table}

Corollary \ref{coroconalg} shows that the approximation in \eqref{convereg1eq2} is more accurate when the sample size is larger. We conduct numerical studies using sample sizes $20,50,100,150,200$. We choose $\theta = 3$. The results are presented in Table \ref{Tab:sec1simunchange}.

\begin{table}[h]
\centering
\begin{tabular}{c|c|c|c}
\hline
Sample size & Regression coefficient & Iteration numbers & Absolute difference\\
\hline
20 & -0.110  & 225.05 & 0.269 \\
50 & -0.410  & 60.26 & 0.0315  \\
100 & -0.404    & 61 & 0.0260 \\
150 & -0.363   & 68 & 0.0148 \\
200 & -0.381  & 65 & 0.00244\\
\hline
\end{tabular}
\caption{The simulation results under different sample sizes. The last column shows the absolute difference between $2\log(\psi(3))$ and regression coefficients, given by $|2\log(\psi(\theta))-$ Regression coefficient$|$.}
\label{Tab:sec1simunchange}
\end{table}

From Tables \ref{Tab:simuResults} and \ref{Tab:sec1simunchange}, we find that the absolute difference increases as $\theta$ increases and sample size decreases. When $\psi(\theta)$ decreases, the iteration number decreases, which implies the convergence of Algorithm \ref{alg:itedou} becomes faster. These results corroborate our theory. The regression coefficients are close to our theoretical assertion $2\log(\psi(\theta))$, which verifies the convergence rate in Theorem \ref{cons11} is sharp.


\subsection{Prediction of double penalty model}\label{secnumconsalg}
To study the prediction performance of double penalty model, we consider two examples, with $L_2$-separable function classes and non-$L_2$-separable function classes, respectively.



\begin{example}
	Consider function \citep{gramacy2012cases}
	\begin{align*}
	h(x) = \frac{\sin(10\pi x)}{2x} + (x-1)^4, x\in[0.5,2.5].
	\end{align*}
	Let $\mathcal{F} =\{f(x) = \beta_1 x + \beta_2, \beta_1^2 + \beta_2^2 \leq 100\}$, and $\mathcal{G}$ be the reproducing kernel Hilbert space generated by the projected kernel. The projected kernel is calculated as in \eqref{projectkdef}, where $\Psi$ is as in \eqref{kernelfunctionPsi} with $\nu = 3.5$ and $\phi = 1$. We use 20 uniformly distributed points from $[0.5,2.5]$ as training points, and let $\epsilon\sim N(0,0.1)$. For each simulation, we calculate the mean squared prediction error, which is approximated by calculating the mean squared prediction error on 201 evenly spaced points. We run 100 simulations, and the average mean squared prediction error is 0.016. In this example, the iteration number needed in Algorithm \ref{alg:itedou} is less than three because the two function classes are orthogonal, which corroborates the results in Corollary \ref{coroconalg}. Figure \ref{simple1d} in Appendix illustrates one simulation result.
\end{example}

%
%



\begin{example}\label{eg2num}
	Consider a modified function of \cite{sun2014balancing} $$h(x) = \frac{2}{\sqrt{\sum_{i=1}^5(x_i - 0.5)^2 } + 1} + \frac{0.5}{\sqrt{\sum_{i=1}^5(x_i - 0.7)^2} + 1},x_i\in [0,1].$$ We use $\mathcal{F} =\{f(x) = \beta_1^T x + \beta_2, \|\beta_1\|_2^2 + \beta_2^2 \leq 10000, x\in[0,1]^5\}$, and $\mathcal{G}$ as the reproducing kernel Hilbert space generated by $\Psi$, where $\Psi$ is as in \eqref{kernelfunctionPsi} with $\nu = 3.5$ and $\phi = 1$. Note that  $\mathcal{F} $ and $\mathcal{G}$ are not $L_2$-separable because $\mathcal{F} \subset \mathcal{G}$.
	
	The double penalty model is
	\begin{align}\label{num5dobj}
	(\hat \beta, \hat g) = \argmin_{\beta \in \mathcal{F}, g\in \mathcal{N}_{\Psi}([0,1]^5)} \|y - x^T\beta - g\|_n^2 + \lambda \|g\|_{\mathcal{N}_{\Psi}([0,1]^5)}^2.
	\end{align}
	We choose $n\lambda = 1, 0.1, 0.01$, where $n=50$ is the sample size. The noise $\epsilon\sim N(0,\sigma^2)$, where $\sigma^2$ is chosen to be $0.1$ and $0.01$. The iteration numbers are fixed in each simulation, with values $1,2,3,4,5$. We choose maximin Latin hypercube
	design \citep{santner2013design} with sample size 50 as the training set. We run 100 simulations for each case and calculated the mean squared prediction error on the testing set, which is the first 1000 points of the Halton sequence \citep{niederreiter1992random}.
	
	For the conciseness of this paper, the simulation results are reported in Appendix. Here we present the main findings from the simulation results: (i) The prediction error in all cases are small, which suggests that the double penalty model can make accurate prediction. (ii) If we increase $n\lambda$, the training error decreases. The prediction error decreases when $n\lambda$ is relatively large, and becomes large when $n\lambda$ is too small. (iii) One iteration in Algorithm \ref{alg:itedou} is sufficient to obtain a good solution of \eqref{num5dobj}. (iv) The training error of the case with smaller $\sigma^2$ is smaller. If $n\lambda$ is chosen properly, the prediction error of the case with smaller $\sigma^2$ is small. However, there is no much difference of the prediction error under the cases $\sigma^2=0.1$ and $0.01$ when $n\lambda$ is large. (v) For all values of $n\lambda$, the $L_2$ norm of the linear function $\hat f$ does not varies a lot. The $L_2$ norm of $\hat g$, on the other hand, increases as $n\lambda$ decreases. This is because a smaller $n\lambda$ implies less penalty on $g$.  (vi) Comparing the values of the $L_2$ norm of $\hat f$ and the $L_2$ norm of $\hat g$, we can see the $L_2$ norm of $\hat f$ is much larger, which is desired because we tend to maximize the interpretable part, which is linear functions in this example.
\end{example}

\section{Application to real datasets}\label{secapptomic}
To illustrate, we apply the approach to two datasets. The first dataset is \cite{goodrich2014human}, which includes 50 human fecal microbiome features for $n=414$ unrelated individuals, of genetic sequence tags corresponding to bacterial taxa, and with a response variable of log-transformed body mass index (BMI). To increase the prediction accuracy, we first reduce the number of original features to the final dataset using the HFE cross-validated approach \citep{oudah2018taxonomy}, as discussed in \cite{zhou2019review}. The second dataset is the diabetes dataset from the {\tt lars} R package, widely used to illustrate penalized regression \citep{hastie2005elements}. The response is a log-transformed measure of disease progression one year after baseline, and predictor features are ten baseline variables, age, sex, BMI, average blood pressure, and six blood serum measurements.

Following Algorithm \ref{alg:itedou}, we let $f$ denote the LASSO algorithm (\cite{tibshirani1996regression}, interpretable part), and use the built-in $l_1$ penalty as $L_f$, with parameter $\lambda_f$ as implemented in the R package {\tt glmnet}. For the ``uninterpretable" part, we use the xgboost decision tree approach, with built-in $L_2$ penalty as $L_g$, with parameter $\lambda_g$ as implemented in the R package {\tt xgboost} \citep{chen2016xgboost}. For xgboost, we set an $L_1$ penalty as zero throughout, with other parameters (tree depth, etc.), set by cross-validation internally, while preserving convexity of $L_g$.  We also set the maximum number of boosting iterations at ten.  At each iterative step of LASSO and xgboost, ten simulations of five-fold cross-validation were performed and the predicted values were then averaged.

Finally, in order to explore the tradeoffs between the interpretable and uninterpretable parts, we first establish a range-finding exercise for the penalty tuning parameters on the logarithmic scale, such that $\log_{10}(\lambda_g)+ \log_{10}(\lambda_f)=c$ for constant $c$. We refer to this tradeoff as the {\it transect} between the tuning parameters, with low values of $\lambda_f$, for example, emphasizing and placing weight on the interpretable part by enforcing a low penalty for overfitting. To illustrate performance, we use the Pearson correlation coefficient between the response vector $y$ and the average (cross-validated) values of $\hat{f}$, $\hat{g}$ and $(\hat{f}+\hat{g})$ over the transect. The correlations are of course directly related to the objective function term $\sum (y-\hat{f}-\hat{g})^2$, but are easier to interpret. Note that $\hat{f}$ and $\hat{g}$ are not orthogonal, so the correlations do not partition into the overall correlation of $y$ with $(\hat{f}+\hat{g})$. Additionally, as a final comparison, we compute these correlation values over the entire grid of $\{\lambda_f,\lambda_g\}$ values, to ensure that the transect was largely capturing the best choice of tuning parameters.

\begin{figure}[h!]
    \centering
    \includegraphics[width=0.5\textwidth]{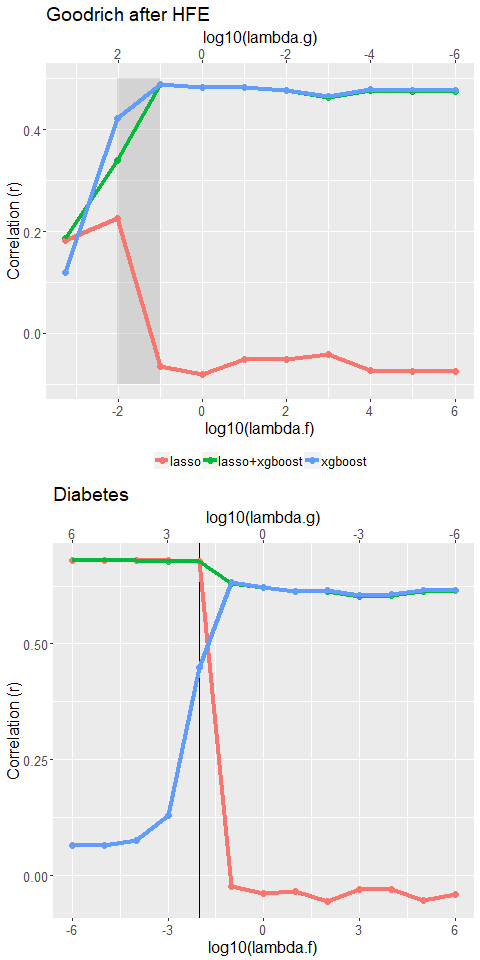}
    \caption{Top panel: cross-validated correlations between $y$ and each of $\hat f$, $\hat g$, and $(\hat{f}+\hat{g})$ for the microbiome dataset, where the tuning parameters vary along the transect as described in the text. Bottom panel: the analogous correlations for the diabetes dataset. Grey region and black vertical line represent suggested tuning parameter values to maximize interpretability while preserving high prediction accuracy.}\label{lasso_xgboost}
\end{figure}

For the Goodrich microbiome data, Figure \ref{lasso_xgboost} top panel shows the correlations between $y$ and the three cross-validated predictors over the transect.  Low values of $\lambda_f$ are favored, although it is clear that the decision tree is favored throughout most of the transect, i.e. $y$ has much higher correlations with $\hat{g}$ than with $\hat{f}$. Using $\log_{10}(\lambda_f)$ in the range of (-2,-1) maximizes the correlation with the interpretable portion, while still achieving near the overall maximum correlation for the combined prediction rule (correlation of nearly 0.5). Our subjective ``best balance" region for the interpretable portion is shown on the figure.
Figure \ref{lasso_xgboost} bottom panel shows the analogous results for the diabetes dataset. Here LASSO provides overall good predictions for small tuning parameter $\lambda_f$, and $\log_{10}(\lambda_f)=-2$ provides good correlations (in the range 0.55-0.6) of $y$ with  $\hat{f}$, $\hat{g}$ and $(\hat{f}+\hat{g})$. As the tuning parameter $\lambda_f$ increases, the correlation between $y$ and $\hat{f}$ falls off dramatically, and our suggested ``best balance" point is also shown. In no instance were the correlation values for the full grid of $\{\lambda_f,\lambda_g\}$ more than 0.015 greater than the greatest value observed along the transects.

\section{Discussion}\label{secconclusion}
In this work, we propose using a double penalty model as a means of balancing tradeoffs between the interpretable part and the non-interpretable portions of a prediction model. Unlike much statistical modelling which focuses primarily on prediction accuracy, we consider the interpretability as an explicit feature. The user can specifies the function classes for the two components of the model, and is able to judge the tradeoff between interpretability and overall prediction accuracy. Our model is very flexible and can be used in many practical scenarios. If two function classes are orthogonal, the convergence of the algorithm provided in this work is very fast. This observation inspires potential future work, given any two function classes, to construct two separable function classes that are orthogonal, and to obtain consistency results, since the two portions are identifiable.

Although our interest here is theoretical, we have also illustrated how the fitting algorithm can be used in practice to make the relative contribution of $\hat{f}$ large, while not substantially degrading overall predictive performance. Further practical implications and implementation issues will be described elsewhere.

\appendix
\numberwithin{equation}{section}
\numberwithin{lemma}{section}
\numberwithin{assumption}{section}

\section{Proof of Lemmas}
\subsection{Proof of Lemma \ref{lemfstarunique}}
The proof is straightforward. Suppose there exist another two functions $f_0\in \mathcal{F}$ and $g_0\in \mathcal{G}$ such that $h = f_0 + g_0$. By \eqref{ineqsepecond}, we have
\newpage
\begin{align*}
    0 & = \|f_0+g_0 - f^* - g^*\|_{L_2}^2\\
    & = \|f_0- f^*\|_{L_2}^2 + \|g_0 - g^*\|_{L_2}^2 + 2\langle f_0- f^*,g_0 - g^*\rangle_2\\
    & \geqslant \|f_0- f^*\|_{L_2}^2 + \|g_0 - g^*\|_{L_2}^2 - 2\theta_1\|f_0- f^*\|_{L_2}\|g_0 - g^*\|_{L_2}\\
    & \geqslant \|f_0- f^*\|_{L_2}^2 + \|g_0 - g^*\|_{L_2}^2 - 2\|f_0- f^*\|_{L_2}\|g_0 - g^*\|_{L_2}\\
    & = (\|f_0- f^*\|_{L_2} - \|g_0 - g^*\|_{L_2})^2,
\end{align*}
where the equality holds only when $\|f_0- f^*\|_{L_2} = \|g_0 - g^*\|_{L_2} = 0$. Thus, we finish the proof.

\subsection{Proof of Lemma \ref{sepeequi}}
The proof of Lemma \ref{sepeequi} relies on the entropy number. Let $(T,d)$ be a metric space with metric $d$, and $T$ is a space. The $\epsilon$-covering number of the metric space $(T,d)$, denoted as $N(\epsilon,T,d)$, is the minimum integer $N$ so that there exist $N$ distinct balls in $(T,d)$ with radius $\epsilon$, and the union of these balls covers $T$. Let $H(\epsilon,T,d) = \log N(\epsilon,T,d)$ be the entropy number. We need the following two lemmas. Lemma \ref{thm:thm21inGeer2014} is a direct result of Theorem 2.1 of \cite{van2014uniform}, which provides an upper bound on the difference between the empirical norm and $L_2$ norm. Lemma \ref{lemmthm31ingeer} is a direct result of Theorem 3.1 of \cite{van2014uniform}, which provides an upper bound on the empirical inner product. In Lemma \ref{thm:thm21inGeer2014} and Lemma \ref{lemmthm31ingeer}, we use the following definition. For $z>0$, we define
\begin{align*}
    J_\infty^2(z,\mathcal{A}) = C_0^2\inf_{\delta>0} \mathbb{E}\left[z\int_\delta^1 \sqrt{H(uz/2,\mathcal{A},\|\cdot\|_\infty)} + \sqrt{n}\delta z \right]^2,
\end{align*}
where $C_0$ is a constant, and $H(u,\mathcal{A},\|\cdot\|_\infty)$ is the entropy of $(\mathcal{A},\|\cdot\|_\infty)$ for a function class $\mathcal{A}$. Furthermore, we assume the following.
\begin{assumption}\label{assumentrf}
$J_\infty(K_1,\mathcal{F})$ and $J_\infty(K_2,\mathcal{G})$ are finite, where $K_1 = \sup_{f\in \mathcal{F}}\|f\|_\infty$ and $K_2 = \sup_{g\in \mathcal{G}}\|g\|_\infty$.
\end{assumption}

\begin{lemma}\label{thm:thm21inGeer2014}
Let $R=\sup_{f\in\mathcal{A}}\|f\|_2$, and $K=\sup_{f\in\mathcal{A}}\|f\|_\infty$, where $\mathcal{A}$ is a class. Then for all $t>0$, with probability at least $1-\exp(-t)$,
\begin{align*}
\sup_{f\in\mathcal{A}}\bigg|\|f\|^2_n-\|f\|^2_2\bigg|\leq C_1\bigg(\frac{2R J_\infty(K,\mathcal{A})+RK\sqrt{t}}{\sqrt{n}}+\frac{4 J_\infty^2(K,\mathcal{A})+K^2t}{n}\bigg),
\end{align*}
where $C_1$ is a constant.
\end{lemma}

\begin{lemma}\label{lemmthm31ingeer}
Let $\mathcal{F}$ and $\mathcal{G}$ be two function classes. Let
\begin{align*}
    R_1 = \sup_{f\in \mathcal{F}}\|f\|_2, K_1 = \sup_{f\in \mathcal{F}}\|f\|_\infty,
\end{align*}
and
\begin{align*}
    R_2 = \sup_{g\in \mathcal{G}}\|g\|_2, K_2 = \sup_{g\in \mathcal{G}}\|g\|_\infty.
\end{align*}
Suppose that $R_1K_2 \leq R_2K_1$. Assume
\begin{align*}
    \left(\frac{2R_1 J_\infty(K_1,\mathcal{F})+ R_1K_1\sqrt{t}}{\sqrt{n}} + \frac{4J_\infty^2(K_1,\mathcal{F}) + K_1^2t}{n}\right)\leq \frac{R_1^2}{C_1},
\end{align*}
and
\begin{align*}
    \left(\frac{2R_2 J_\infty(K_2,\mathcal{G})+ R_2K_2\sqrt{t}}{\sqrt{n}} + \frac{4J_\infty^2(K_2,\mathcal{G}) + K_2^2t}{n}\right)\leq \frac{R_2^2}{C_1}.
\end{align*}
Then for $t\geqslant 4$, with probability at least $1-12\exp(-t)$,
\begin{align*}
    \frac{1}{8C_1}\sup_{f\in \mathcal{F},g\in \mathcal{G}}\left|\langle f,g\rangle_2 - \langle f,g\rangle_n\right|\leq \frac{R_1J_\infty(K_2,\mathcal{G}) + R_2J_\infty(R_1K_2/R_2),\mathcal{F}) +R_1K_2\sqrt{t}}{\sqrt{n}} + \frac{K_1K_2t}{n}.
\end{align*}
\end{lemma}
Now we are ready to prove Lemma \ref{sepeequi}. This Lemma is a direct result of Lemma \ref{thm:thm21inGeer2014} and Lemma \ref{lemmthm31ingeer}. To see this, suppose
\begin{align}\label{pflemcond2}
    |\langle f,g\rangle_2| \leq \theta_1\|f\|_2\|g\|_2,
\end{align}
for all $f\in \mathcal{F}$ and $g\in\mathcal{G}$. By Lemma \ref{thm:thm21inGeer2014}, for
\newline
$t\in [\max \{2J_\infty(K_1,\mathcal{F})/K_1,2J_\infty(K_2,\mathcal{G})/K_2\},\min \{nR_1^2/K_1^2, nR_2^2/K_2^2\}]$, and $n$ sufficient large such that $\max \{2J_\infty(K_1,\mathcal{F})/K_1,2J_\infty(K_2,\mathcal{G})/K_2\}\leq\min \{nR_1^2/K_1^2, nR_2^2/K_2^2\}$,
\begin{align*}
    \|f\|_2^2 \leq \|f\|_n^2 + C_1 \frac{\sqrt{t}}{\sqrt{n}}, \text{ and } \|g\|_2^2 \leq \|g\|_n^2 + C_1 \frac{\sqrt{t}}{\sqrt{n}},
\end{align*}
which implies
\begin{align}\label{pflemnnorm2}
    \|f\|_2^2 \|g\|_2^2 & \leq \bigg(\|f\|_n^2 + C_1 \frac{\sqrt{t}}{\sqrt{n}}\bigg)\bigg(\|g\|_n^2 + C_1 \frac{\sqrt{t}}{\sqrt{n}}\bigg)\nonumber\\
    & = \|f\|_n^2\|g\|_n^2 + C_1 \frac{\sqrt{t}}{\sqrt{n}}(\|f\|_n^2 + \|g\|_n^2) + C_1^2\frac{t}{n}\nonumber\\
    & \leq  \|f\|_n^2\|g\|_n^2 + C_1(K_1^2+K_2^2) \frac{\sqrt{t}}{\sqrt{n}} + C_1^2\frac{t}{n}.
\end{align}
For $t\leq n(K_1^2+K_2^2)^2/C_1^2$, \eqref{pflemnnorm2} implies
\begin{align*}
    \|f\|_2^2 \|g\|_2^2 & \leq   \|f\|_n^2\|g\|_n^2 + C_2\frac{\sqrt{t}}{\sqrt{n}}.
\end{align*}
By using elementary inequality $\sqrt{a+b} \leq \sqrt{a} + \sqrt{b}$, we have
\begin{align}\label{pflemnnorm3}
    \|f\|_2 \|g\|_2 & \leq   \|f\|_n\|g\|_n + C_3\frac{t^{1/4}}{n^{1/4}}.
\end{align}
Similarly, by Lemma \ref{lemmthm31ingeer}, we have
\begin{align}\label{pflemninner2}
    |\langle f,g\rangle_2|\geqslant |\langle f,g\rangle_n| -  C_4 \frac{\sqrt{t}}{\sqrt{n}}.
\end{align}
Plugging \eqref{pflemnnorm3}, \eqref{pflemninner2} into \eqref{pflemcond2} yields
\begin{align*}
    |\langle f,g\rangle_n| \leq \theta_1\|f\|_n\|g\|_n + C_4 \frac{\sqrt{t}}{\sqrt{n}} + C_3\frac{t^{1/4}}{n^{1/4}} \leq \theta_1\|f\|_n\|g\|_n + C_5\frac{t^{1/4}}{n^{1/4}},
\end{align*}
for $t\leq nC_3^4/C_4^4$. It can be seen either $\|f\|_n\|g\|_n \leq C_5t^{1/4}n^{-1/4 + \alpha}$ or $|\langle f,g\rangle_n| \leq (\theta_1 + C_6n^{-\alpha}) \|f\|_n\|g\|_n$, which finishes the proof.

\section{Proof of Theorem \ref{cons11}}

If $m=1$, then the results automatically hold. Suppose $m>1$. Since $f_m$ is the solution to \eqref{fmstep}, for any $\alpha \in (0,1)$, we have
\begin{align}\label{pfthmeq1}
    & \|f^* + g^* + \epsilon - f_m - g_m\|_n^2 + L_f(f_m)\nonumber\\
    \leq & \|f^* + g^* + \epsilon - \alpha \hat f - (1-\alpha)f_m - g_m\|_n^2 + L_f(\alpha \hat f + (1-\alpha)f_m)\nonumber\\
   \leq & \|f^* + g^* + \epsilon - \alpha \hat f - (1-\alpha)f_m - g_m\|_n^2 + \alpha L_f(\hat f) + (1-\alpha)L_f(f_m),
\end{align}
where the last inequality is because $L_f$ is convex. Rewriting \eqref{pfthmeq1} yields
\begin{align*}
    & \|f^* - f_m\|_n^2 + 2\langle f^* - f_m, g^* - g_m + \epsilon \rangle_n + L_f(f_m)\\
    \leq & \alpha^2 \|f^* - \hat f\|_n^2 + (1-\alpha)^2\|f^* - f_m\|_n^2 + 2\langle f^* - \alpha \hat f - (1-\alpha)f_m, g^* - g_m + \epsilon \rangle\\
    & + 2\alpha(1-\alpha)\langle f^* - \hat f, f^* - f_m \rangle_n + \alpha L_f(\hat f) + (1-\alpha)L_f(f_m),
\end{align*}
which is the same as
\begin{align}\label{thm1equba2}
    & (2\alpha - \alpha^2)\|f^* - f_m\|_n^2 +  2\alpha\langle f^* - f_m, g^* - g_m + \epsilon \rangle_n + \alpha L_f(f_m)\nonumber\\
    \leq & \alpha^2 \|f^* - \hat f\|_n^2 + 2\alpha\langle f^* - \hat f, g^* - g_m + \epsilon \rangle_n + 2\alpha(1-\alpha)\langle f^* - \hat f, f^* - f_m \rangle_n + \alpha L_f(\hat f).
\end{align}
Because $\alpha \in (0,1)$, \eqref{thm1equba2} implies
\begin{align}\label{thm1equba3}
    & (2 - \alpha)\|f^* - f_m\|_n^2 +  2\langle f^* - f_m, g^* - g_m + \epsilon \rangle_n +  L_f(f_m)\nonumber\\
    \leq & \alpha \|f^* - \hat f\|_n^2 + 2\langle f^* - \hat f, g^* - g_m + \epsilon \rangle_n + 2(1-\alpha)\langle f^* - \hat f, f^* - f_m \rangle_n +  L_f(\hat f).
\end{align}
Taking limit $\alpha \rightarrow 0$ in \eqref{thm1equba3} leads to
\begin{align}\label{thm1equba4}
    & \|f^* - f_m\|_n^2 +  \langle f^* - f_m, g^* - g_m + \epsilon \rangle_n +  L_f(f_m)/2\nonumber\\
    \leq & \langle f^* - \hat f, g^* - g_m + \epsilon \rangle_n + \langle f^* - \hat f, f^* - f_m \rangle_n +  L_f(\hat f)/2.
\end{align}

Since $\hat f$ is the solution to \eqref{ourPMmodel}, for any $\beta \in (0,1)$, it is true that
\begin{align*}
    & \|f^* + g^* + \epsilon - \hat f - \hat g\|_n^2 + L_f(\hat f) + L_g(\hat g) \nonumber\\
    \leq & \|f^* + g^* + \epsilon - \beta \hat f - (1-\beta) f_m - \hat g\|_n^2 + L_f( \beta \hat f + (1-\beta) f_m ) + L_g(\hat g)\\
    \leq &  \|f^* + g^* + \epsilon - \beta \hat f - (1-\beta) f_m - \hat g\|_n^2 + \beta L_f( \hat f) + (1-\beta) L_f( f_m ) + L_g(\hat g),
\end{align*}
which implies
\begin{align}\label{thm1equbbeta1}
   & (1 - \beta^2)\|f^* - \hat f\|_n^2 + 2(1-\beta)\langle f^* - \hat f, g^* - \hat g + \epsilon \rangle_n + (1-\beta)L_f(\hat f)\nonumber\\
   \leq & (1-\beta)^2\|f^* - f_m\|_n^2 + 2\beta(1-\beta)\langle f^* - \hat f, f^* - f_m \rangle_n \nonumber\\
   & + 2(1-\beta)\langle f^* - f_m, g^* - \hat g + \epsilon \rangle_n + (1-\beta) L_f( f_m ).
\end{align}
Since $\beta < 1$, \eqref{thm1equbbeta1} implies
\begin{align}\label{thm1equbbeta2}
   & (1 + \beta)\|f^* - \hat f\|_n^2 + 2\langle f^* - \hat f, g^* - \hat g + \epsilon \rangle_n + L_f(\hat f)\nonumber\\
   \leq & (1-\beta)\|f^* - f_m\|_n^2 + 2\beta\langle f^* - \hat f, f^* - f_m \rangle_n \nonumber\\
   & + 2\langle f^* - f_m, g^* - \hat g + \epsilon \rangle_n + L_f( f_m ).
\end{align}
Letting $\beta\rightarrow 1$ in \eqref{thm1equbbeta2} yields
\begin{align}\label{thm1equbbeta3}
   & \|f^* - \hat f\|_n^2 + \langle f^* - \hat f, g^* - \hat g + \epsilon \rangle_n + L_f(\hat f)/2\nonumber\\
   \leq & \langle f^* - \hat f, f^* - f_m \rangle_n  + \langle f^* - f_m, g^* - \hat g + \epsilon \rangle_n + L_f( f_m )/2.
\end{align}
Combining \eqref{thm1equbbeta3} and \eqref{thm1equba4}, it can be checked that
\begin{align}\label{pfthm1ffinial}
    \|\hat f - f_m\|_n^2 \leq -\langle \hat f - f_m, \hat g - g_m \rangle_n.
\end{align}
By the separability of function classes $\mathcal{F}$ and $\mathcal{G}$ with respect to the empirical norm, \eqref{pfthm1ffinial} implies
\begin{align}\label{pfthm1ffinial2}
    & \|\hat f - f_m\|_n^2 \leq \theta_1 \|\hat f - f_m\|_n\|\hat g - g_m\|_n\nonumber\\
    \Leftrightarrow & \|\hat f - f_m\|_n \leq \theta_1\|\hat g - g_m\|_n.
\end{align}
Applying the same procedure to function $g_{m+1}$, we have
\begin{align}\label{pfthm1gfinial2}
    \|\hat g - g_{m+1}\|_n \leq & \theta_1 \|\hat f - f_m\|_n.
\end{align}
By \eqref{pfthm1ffinial2} and \eqref{pfthm1gfinial2}, it can be seen that
\begin{align*}
    \|\hat g - g_{m+1}\|_n \leq & \theta_1 \|\hat f - f_m\|_n\\
    \leq & \theta_1^2 \|\hat g - g_m\|_n...\leq \theta_1^{2m-2} \|\hat g - g_1\|_n,
\end{align*}
which converges to zero, since $\theta_1 < 1$. Similarly, $\|\hat f - f_m\|_n$ converges to zero.

By \eqref{ourPMmodel} and \eqref{gmstep},
\begin{align}\label{Lgmless}
    \|y_i - f_{m-1} - g_m\|_n^2 + L_g(g_m) \leq\|y_i - f_{m-1}  - \hat g\|_n^2 + L_g(\hat g),
\end{align}
and
\begin{align}\label{Lghatless}
    \|y_i - \hat f - \hat g\|_n^2 + L_g(\hat g) \leq \|y_i - \hat f - g_m\|_n^2 + L_g(g_m).
\end{align}
Since $\|\hat f - f_m\|_n$ and $\|\hat g - g_m\|_n$ converges to zero, we have for each $x_i$, $f_m(x_i)$ and $g_m(x_i)$ converges to $f(x_i)$ and $g(x_i)$, respectively. By \eqref{Lgmless} and \eqref{Lghatless}, we have $L_g(g_m)$ converges to $L_g(\hat g)$. Similarly, $L_f(f_m)$ converges to $L_f(\hat f)$.

\section{Proof of Corollary \ref{coroconalg}}

The proof of Corollary \ref{coroconalg} is similar to the proof of Theorem \ref{cons11}. We only need to note that if $\|\hat f - f_m\|_n \|\hat g - g_m\|_n \leq C_5t^{1/4}n^{-1/4 + \alpha}$, then by \eqref{pfthm1ffinial}, we have
\begin{align*}
    \|\hat f - f_m\|_n^2 \leq -\langle \hat f - f_m, \hat g - g_m \rangle_n\leq \|\hat f - f_m\|_n \|\hat g - g_m\|_n \leq C_5t^{1/4}n^{-1/4 + \alpha}.
\end{align*}
Similarly, we have
\begin{align*}
    \|\hat g - g_{m+1}\|_n^2 \leq & \|\hat f - f_m\|_n^2 \leq C_5t^{1/4}n^{-1/4 + \alpha}.
\end{align*}

If $\|\hat f - f_m\|_n \|\hat g - g_m\|_n > C_5t^{1/4}n^{-1/4 + \alpha}$, then the results follow from Lemma \ref{sepeequi}.

\section{Proof of Theorem \ref{cons22}}
Without loss of generality, assume $L_f$ is strongly convex. For any $\alpha \in (0,1)$, by the strong convexity of $L_f$, we have
\begin{align}\label{pfthmeq1sconv}
    & \|f^* + g^* + \epsilon - f_m - g_m\|_n^2 + L_f(f_m)\nonumber\\
    \leq & \|f^* + g^* + \epsilon - \alpha \hat f - (1-\alpha)f_m - g_m\|_n^2 + L_f(\alpha \hat f + (1-\alpha)f_m)\nonumber\\
   \leq & \|f^* + g^* + \epsilon - \alpha \hat f - (1-\alpha)f_m - g_m\|_n^2 + \alpha L_f(\hat f) + (1-\alpha)L_f(f_m)-\frac{1}{2}\gamma \alpha(1-\alpha)\|\hat f-f_m\|_n^2.
\end{align}
Similar to the proof of Theorem \ref{cons11}, we can rewrite \eqref{pfthmeq1sconv} as
\begin{align*}
    & \|f^* - f_m\|_n^2 + 2\langle f^* - f_m, g^* - g_m + \epsilon \rangle_n + L_f(f_m)\\
    \leq & \alpha^2 \|f^* - \hat f\|_n^2 + (1-\alpha)^2\|f^* - f_m\|_n^2 + 2\langle f^* - \alpha \hat f - (1-\alpha)f_m, g^* - g_m + \epsilon \rangle\\
    & + 2\alpha(1-\alpha)\langle f^* - \hat f, f^* - f_m \rangle_n + \alpha L_f(\hat f) + (1-\alpha)L_f(f_m)-\frac{1}{2}\gamma \alpha(1-\alpha)\|\hat f-f_m\|_n^2\\
    \leq & \alpha^2 \|f^* - \hat f\|_n^2 + (1-\alpha)^2\|f^* - f_m\|_n^2 + 2\langle f^* - \alpha \hat f - (1-\alpha)f_m, g^* - g_m + \epsilon \rangle\\
    & + 2\alpha(1-\alpha)\langle f^* - \hat f, f^* - f_m \rangle_n + \alpha L_f(\hat f) + (1-\alpha)L_f(f_m)\\
    & - \frac{1}{2}\gamma \alpha(1-\alpha)(\|\hat f - f^*\|_n^2 - 2\langle f^* - \hat f, f^* - f_m \rangle_n  + \|f^* -f_m\|_n^2),
\end{align*}
which is the same as
\begin{align}\label{thm1equba2sconv}
    & (2\alpha - \alpha^2 + \frac{1}{2}\gamma \alpha(1-\alpha))\|f^* - f_m\|_n^2 +  2\alpha\langle f^* - f_m, g^* - g_m + \epsilon \rangle_n + \alpha L_f(f_m)\nonumber\\
    \leq & (\alpha^2 - \frac{1}{2}\gamma \alpha(1-\alpha)) \|f^* - \hat f\|_n^2 + 2\alpha\langle f^* - \hat f, g^* - g_m + \epsilon \rangle_n\nonumber\\
    & + (2 + \gamma)\alpha(1-\alpha)\langle f^* - \hat f, f^* - f_m \rangle_n + \alpha L_f(\hat f)\nonumber\\
    \Leftrightarrow & (2 - \alpha + \frac{1}{2}\gamma (1-\alpha))\|f^* - f_m\|_n^2 +  2\langle f^* - f_m, g^* - g_m + \epsilon \rangle_n + L_f(f_m)\nonumber\\
    \leq & (\alpha - \frac{1}{2}\gamma (1-\alpha)) \|f^* - \hat f\|_n^2 + 2\langle f^* - \hat f, g^* - g_m + \epsilon \rangle_n\nonumber\\
    & + (2 + \gamma)(1-\alpha)\langle f^* - \hat f, f^* - f_m \rangle_n + L_f(\hat f).
\end{align}
Taking limit $\alpha \rightarrow 0$ in \eqref{thm1equba2sconv} yields
\begin{align}\label{thm1equba4sconv}
    & (2 + \frac{1}{2}\gamma )\|f^* - f_m\|_n^2 +  2\langle f^* - f_m, g^* - g_m + \epsilon \rangle_n + L_f(f_m)\nonumber\\
    \leq &  - \frac{1}{2}\gamma \|f^* - \hat f\|_n^2 + 2\langle f^* - \hat f, g^* - g_m + \epsilon \rangle_n + (2 + \gamma)\langle f^* - \hat f, f^* - f_m \rangle_n + L_f(\hat f).
\end{align}
Since $\hat f$ is the solution to \eqref{ourPMmodel}, for any $\beta \in (0,1)$, it is true that
\begin{align*}
    & \|f^* + g^* + \epsilon - \hat f - \hat g\|_n^2 + L_f(\hat f) + L_g(\hat g) \nonumber\\
    \leq & \|f^* + g^* + \epsilon - \beta \hat f - (1-\beta) f_m - \hat g\|_n^2 + L_f( \beta \hat f + (1-\beta) f_m ) + L_g(\hat g)\\
    \leq &  \|f^* + g^* + \epsilon - \beta \hat f - (1-\beta) f_m - \hat g\|_n^2 + \beta L_f( \hat f) + (1-\beta) L_f( f_m ) + L_g(\hat g) -\frac{1}{2}\gamma \beta(1-\beta)\|\hat f-f_m\|_n^2.
\end{align*}
By the similar approach as shown in \eqref{pfthmeq1sconv} - \eqref{thm1equba4sconv}, we can show
\begin{align}\label{thm1equba44sconv}
    & (2 + \frac{1}{2}\gamma )\|f^* - \hat f\|_n^2 +  2\langle f^* - \hat f, g^* - \hat g + \epsilon \rangle_n + L_f(\hat f)\nonumber\\
    \leq &  - \frac{1}{2}\gamma \|f^* - f_m\|_n^2 + 2\langle f^* - f_m, g^* - \hat g + \epsilon \rangle_n + (2 + \gamma)\langle f^* - \hat f, f^* - f_m \rangle_n + L_f(f_m).
\end{align}
Combining \eqref{thm1equba4sconv} and \eqref{thm1equba44sconv} leads to
\begin{align*}
(1 + \frac{1}{2}\gamma )\|\hat f - f_m\|_n^2 \leq -\langle \hat f - f_m, \hat g - g_m \rangle_n\leq \|\hat f - f_m\|_n\|\hat g - g_m\|_n.
\end{align*}
Thus,
\begin{align}\label{pfthm1ffinialsconv2}
(1 + \frac{1}{2}\gamma )\|\hat f - f_m\|_n \leq \|\hat g - g_m\|_n.
\end{align}
Applying the same procedure to function $g_{m+1}$, and noting that we do not have the strong convexity of $L_g(g)$, we have
\begin{align}\label{pfthm1gfinial2convg}
    \|\hat g - g_{m+1}\|_n \leq & \|\hat f - f_m\|_n.
\end{align}
By \eqref{pfthm1ffinialsconv2} and \eqref{pfthm1gfinial2convg}, we have
\begin{align*}
    \|\hat g - g_{m+1}\|_n \leq \|\hat f - f_m\|_n \leq \frac{2}{2+\gamma} \|\hat g - g_m\|_n \leq ... \leq \left(\frac{2}{2+\gamma} \right)^m\|\hat g - g_1\|_n,
\end{align*}
which implies $\|\hat g - g_{m}\|_n$ converges to zero. By \eqref{pfthm1ffinialsconv2}, $\|\hat f - f_m\|_n$ also converges to zero. The rest of the proof is similar to the proof of Theorem \ref{cons11}. Thus, we finish the proof.

\section{Proof of Theorem \ref{submodel2thmconsist}}\label{secpfthmprok}
We need the following lemma.

\begin{lemma}[Lemma 8.2 of \cite{geer2000empirical}]\label{bernineqlem}
Let $\epsilon_i$'s be i.i.d. sub-Gaussian random variables, i.e., satisfying $K^2\mathbb{E}\exp(|\epsilon_i|^2/K^2) -1\leq \sigma_0^2$ for some constants $K$ and $\sigma_0^2$, and all $i=1,...,n$. Then for all $\gamma = (\gamma_1,...,\gamma_n)^T\in \RR^n$ and $a>0$,
\begin{align*}
    P\bigg(\bigg|\frac{1}{n}\sum_{k=1}^n \epsilon_k\gamma_k \bigg| \geqslant a\bigg) \leq 2 \exp\bigg[-\frac{n^2a^2}{C\sum_{k=1}^n \gamma_k^2}\bigg],
\end{align*}
where $C$ is a constant depending on $K$ and $\sigma_0^2$.
\end{lemma}
Now we are ready to prove Theorem \ref{submodel2thmconsist}.

Because $\hat \beta$ and $\hat g$ are derived by \eqref{oursubmodel2p},
we have
\begin{align*}
    \|y - x^T\hat \beta - \hat g\|_n^2 + \lambda \|\hat g\|_{\mathcal{N}_{\Psi_{\mathcal{F}}}(\Omega)}^2 \leq \|y - x^T\beta^* - g^*\|_n^2 + \lambda \|g^*\|_{\mathcal{N}_{\Psi_{\mathcal{F}}}(\Omega)}^2,
\end{align*}
which can be rewritten as
\begin{align}\label{Pfthm2basicineq}
& \|x^T(\beta^* - \hat \beta)\|_n^2 + \|g^* - \hat g\|_n^2 + 2\langle x^T(\beta^* - \hat \beta), g^* - \hat g\rangle_n  + \lambda \|\hat g\|_{\mathcal{N}_{\Psi_{\mathcal{F}}}(\Omega)}^2\nonumber\\
\leq &  2\langle \epsilon, x^T(\hat \beta - \beta^*)\rangle_n + 2\langle \epsilon,\hat g - g^*\rangle_n  + \lambda \|g^*\|_{\mathcal{N}_{\Psi_{\mathcal{F}}}(\Omega)}^2.
\end{align}
Note $\mathcal{N}_{\Psi}(\Omega)$ coincides $H^{\nu}(\Omega)$. By the entropy number of a unit ball in the Sobolev space $H^{\nu}(\Omega)$ \citep{adams2003sobolev} and
Lemma 8.4 of \cite{geer2000empirical}, it can be shown that
\begin{align*}
    \sup_{g\in \mathcal{G}}\frac{|\langle \epsilon,\hat g - g^*\rangle_n|}{\|g^* - \hat g\|_n^{1-\frac{p}{2\nu}}(\|\hat g\|_{\mathcal{N}_{\Psi}(\Omega)} + \|g^*\|_{\mathcal{N}_{\Psi}(\Omega)})^{\frac{p}{2\nu}}} = O_P(n^{-1/2}),
\end{align*}
which implies
\begin{align}\label{thm2pfineg}
\langle \epsilon,\hat g - g^*\rangle_n = O_P(n^{-1/2})\|g^* - \hat g\|_n^{1-\frac{p}{2\nu}}(\|\hat g\|_{\mathcal{N}_{\Psi}(\Omega)} + \|g^*\|_{\mathcal{N}_{\Psi}(\Omega)})^{\frac{p}{2\nu}}.
\end{align}
By Lemma \ref{bernineqlem}, we have
\begin{align}\label{thm2pfinebeta}
\langle \epsilon, x^T(\hat \beta - \beta^*)\rangle_n = O_P(n^{-1/2})\|x^T(\beta^* - \hat \beta)\|_n.
\end{align}
Note that $\langle x^T(\beta^* - \hat \beta), g^* - \hat g\rangle_2 = 0$. Therefore, by Lemma \ref{lemmthm31ingeer}
it can be shown that
\begin{align}\label{thm2pfinebetag}
   |\langle x^T(\beta^* - \hat \beta), g^* - \hat g\rangle_n| &  = O_P(n^{-1/2})\|x^T(\beta^* - \hat \beta)\|_{L_\infty}  (\|\hat g\|_{\mathcal{N}_{\Psi}(\Omega)} + \|g^*\|_{\mathcal{N}_{\Psi}(\Omega)}).
\end{align}
By Theorem 3.3 of \cite{tuo2019adjustments}, there exist constants $C_1$ and $C_2$ such that
\begin{align}\label{thm2pfequipkc}
    C_1\|g^*\|_{\mathcal{N}_{\Psi}(\Omega)}^2& \leq \|g^*\|_{\mathcal{N}_{\Psi_{\mathcal{F}}}(\Omega)}^2 \leq C_2\|g^*\|_{\mathcal{N}_{\Psi}(\Omega)}^2,\nonumber\\
    \text{ and }C_1\|\hat g\|_{\mathcal{N}_{\Psi}(\Omega)}^2& \leq \|\hat g\|_{\mathcal{N}_{\Psi_{\mathcal{F}}}(\Omega)}^2 \leq C_2\|\hat g\|_{\mathcal{N}_{\Psi}(\Omega)}^2.
\end{align}
Plugging \eqref{thm2pfineg} - \eqref{thm2pfequipkc} into \eqref{Pfthm2basicineq}, we have
\begin{align}\label{Pfthm2basicineqX}
& \|x^T(\beta^* - \hat \beta)\|_n^2 + \|g^* - \hat g\|_n^2 + \lambda \|\hat g\|_{\mathcal{N}_{\Psi}(\Omega)}^2\nonumber\\
\leq &  O_P(n^{-1/2})\|g^* - \hat g\|_n^{1-\frac{p}{2\nu}}(\|\hat g\|_{\mathcal{N}_{\Psi}(\Omega)} +  \|g^*\|_{\mathcal{N}_{\Psi}(\Omega)})^{\frac{p}{2\nu}} + O_P(n^{-1/2})\|x^T(\beta^* - \hat \beta)\|_n  + \lambda \|g^*\|_{\mathcal{N}_{\Psi}(\Omega)}^2\nonumber\\
& + O_P(n^{-1/2})\|x^T(\beta^* - \hat \beta)\|_{L_\infty}  (\|\hat g\|_{\mathcal{N}_{\Psi}(\Omega)} + \|g^*\|_{\mathcal{N}_{\Psi}(\Omega)}).
\end{align}
Next, we consider two cases

\textbf{Case 1:} $\|\hat g\|_{\mathcal{N}_{\Psi}(\Omega)} \geqslant \|g^*\|_{\mathcal{N}_{\Psi}(\Omega)}$. By \eqref{Pfthm2basicineqX}, we have
\begin{align}\label{Pfthm2hatgda1}
& \|x^T(\beta^* - \hat \beta)\|_n^2 + \|g^* - \hat g\|_n^2 + \lambda \|\hat g\|_{\mathcal{N}_{\Psi}(\Omega)}^2\nonumber\\
\leq &  O_P(n^{-1/2})\|g^* - \hat g\|_n^{1-\frac{p}{2\nu}}\|\hat g\|_{\mathcal{N}_{\Psi}(\Omega)}^{\frac{p}{2\nu}} + O_P(n^{-1/2})\|x^T(\beta^* - \hat \beta)\|_n  + \lambda \|g^*\|_{\mathcal{N}_{\Psi}(\Omega)}^2\nonumber\\
& + O_P(n^{-1/2})\|x^T(\beta^* - \hat \beta)\|_{L_\infty}\|\hat g\|_{\mathcal{N}_{\Psi}(\Omega)}.
\end{align}
By H\"older's inequality and the compactness of $\Omega$,
\begin{align}\label{xbhol}
    \|x^T(\beta^* - \hat \beta)\|_{L_\infty} \leq C_3\|\beta^* - \hat \beta\|_1\leq C_4\|\beta^* - \hat \beta\|_2.
\end{align}
By Hoeffding's inequality,
\begin{align}\label{xbhoe}
    & \|x^T(\beta^* - \hat \beta)\|_n^2 = O_P(\|\beta^* - \hat \beta\|_2^2),\nonumber\\
    \text{and } & \|\beta^* - \hat \beta\|_2^2 = O_P(\|x^T(\beta^* - \hat \beta)\|_n^2).
\end{align}
Plugging \eqref{xbhol} and \eqref{xbhoe} into \eqref{Pfthm2hatgda1}, we have
\begin{align}\label{Pfthm2hatgda2}
& C_5\|\beta^* - \hat \beta\|_2^2 + \|g^* - \hat g\|_n^2 + \lambda \|\hat g\|_{\mathcal{N}_{\Psi}(\Omega)}^2\nonumber\\
\leq &  O_P(n^{-1/2})\|g^* - \hat g\|_n^{1-\frac{p}{2\nu}}\|\hat g\|_{\mathcal{N}_{\Psi}(\Omega)}^{\frac{p}{2\nu}} + O_P(n^{-1/2})\|\beta^* - \hat \beta\|_2  + \lambda \|g^*\|_{\mathcal{N}_{\Psi}(\Omega)}^2\nonumber\\
& + O_P(n^{-1/2})\|\beta^* - \hat \beta\|_2\|\hat g\|_{\mathcal{N}_{\Psi}(\Omega)}.
\end{align}
We consider two subcases.

\textbf{Case 1.1:} $\|\beta^* - \hat \beta\|_2 \leq \|g^* - \hat g\|_n$. By \eqref{Pfthm2hatgda2}, we have
\begin{align*}
& \|g^* - \hat g\|_n^2 + \lambda \|\hat g\|_{\mathcal{N}_{\Psi}(\Omega)}^2\nonumber\\
\leq &  O_P(n^{-1/2})\|g^* - \hat g\|_n^{1-\frac{p}{2\nu}}\|\hat g\|_{\mathcal{N}_{\Psi}(\Omega)}^{\frac{p}{2\nu}} + O_P(n^{-1/2})\|g^* - \hat g\|_n  + \lambda \|g^*\|_{\mathcal{N}_{\Psi}(\Omega)}^2\nonumber\\
& + O_P(n^{-1/2})\|g^* - \hat g\|_n\|\hat g\|_{\mathcal{N}_{\Psi}(\Omega)}\nonumber\\
\leq &  O_P(n^{-1/2})\|g^* - \hat g\|_n^{1-\frac{p}{2\nu}}\|\hat g\|_{\mathcal{N}_{\Psi}(\Omega)}^{\frac{p}{2\nu}}  + \lambda \|g^*\|_{\mathcal{N}_{\Psi}(\Omega)}^2 + O_P(n^{-1/2})\|g^* - \hat g\|_n\|\hat g\|_{\mathcal{N}_{\Psi}(\Omega)}.
\end{align*}
Then either
\begin{align}\label{Pfthm2hatgda2112}
& \|g^* - \hat g\|_n^2 + \lambda \|\hat g\|_{\mathcal{N}_{\Psi}(\Omega)}^2\nonumber\\
\leq &  O_P(n^{-1/2})\|g^* - \hat g\|_n^{1-\frac{p}{2\nu}}\|\hat g\|_{\mathcal{N}_{\Psi}(\Omega)}^{\frac{p}{2\nu}} + O_P(n^{-1/2})\|g^* - \hat g\|_n
+ O_P(n^{-1/2})\|g^* - \hat g\|_n\|\hat g\|_{\mathcal{N}_{\Psi}(\Omega)},
\end{align}
or
\begin{align}\label{Pfthm2hatgda2113}
\|g^* - \hat g\|_n^2 + \lambda \|\hat g\|_{\mathcal{N}_{\Psi}(\Omega)}^2 \leq  4\lambda \|g^*\|_{\mathcal{N}_{\Psi}(\Omega)}^2.
\end{align}
It can be seen that \eqref{Pfthm2hatgda2113} implies
\begin{align*}
    \|\hat g\|_{\mathcal{N}_{\Psi}(\Omega)}^2 & = O_P(1),
    \|g^* - \hat g\|_n^2  = O_P(\lambda),
    \|\beta^* - \hat \beta\|_2^2  = O_P(\lambda).
\end{align*}
Under \eqref{Pfthm2hatgda2112}, we have either
\begin{align}\label{Pfthm2hatgda21121}
\|g^* - \hat g\|_n^2 + \lambda \|\hat g\|_{\mathcal{N}_{\Psi}(\Omega)}^2
\leq  O_P(n^{-1/2})\|g^* - \hat g\|_n^{1-\frac{p}{2\nu}}\|\hat g\|_{\mathcal{N}_{\Psi}(\Omega)}^{\frac{p}{2\nu}},
\end{align}
or
\begin{align}\label{Pfthm2hatgda21122}
\|g^* - \hat g\|_n^2 + \lambda \|\hat g\|_{\mathcal{N}_{\Psi}(\Omega)}^2 \leq O_P(n^{-1/2})\|g^* - \hat g\|_n\|\hat g\|_{\mathcal{N}_{\Psi}(\Omega)}.
\end{align}
Solving \eqref{Pfthm2hatgda21121} leads to
\begin{align*}
    \|\hat g\|_{\mathcal{N}_{\Psi}(\Omega)}^2 & = O_P(1),
    \|g^* - \hat g\|_n^2  = O_{P}(n^{-\frac{2\nu}{2\nu + p}}),
    \|\beta^* - \hat \beta\|_2^2  = O_{P}(n^{-\frac{2\nu}{2\nu + p}}).
\end{align*}
Solving \eqref{Pfthm2hatgda21122} yields $\lambda = O_P(n^{-1})$, which leads to a contradiction.

\textbf{Case 1.2:} $\|\beta^* - \hat \beta\|_2 \geqslant \|g^* - \hat g\|_n$. By \eqref{Pfthm2hatgda2}, we have
\begin{align*}
& C_5\|\beta^* - \hat \beta\|_2^2 + \lambda \|\hat g\|_{\mathcal{N}_{\Psi}(\Omega)}^2\nonumber\\
\leq &  O_P(n^{-1/2})\|\beta^* - \hat \beta\|_2^{1-\frac{p}{2\nu}}\|\hat g\|_{\mathcal{N}_{\Psi}(\Omega)}^{\frac{p}{2\nu}} + O_P(n^{-1/2})\|\beta^* - \hat \beta\|_2  + \lambda \|g^*\|_{\mathcal{N}_{\Psi}(\Omega)}^2\nonumber\\
& + O_P(n^{-1/2})\|\beta^* - \hat \beta\|_2\|\hat g\|_{\mathcal{N}_{\Psi}(\Omega)}\nonumber\\
\leq &  O_P(n^{-1/2})\|\beta^* - \hat \beta\|_2^{1-\frac{p}{2\nu}}\|\hat g\|_{\mathcal{N}_{\Psi}(\Omega)}^{\frac{p}{2\nu}}  + \lambda \|g^*\|_{\mathcal{N}_{\Psi}(\Omega)}^2 + O_P(n^{-1/2})\|\beta^* - \hat \beta\|_2\|\hat g\|_{\mathcal{N}_{\Psi}(\Omega)}.
\end{align*}
By similar approach as in Case 1.1, we have
\begin{align*}
    \|\hat g\|_{\mathcal{N}_{\Psi}(\Omega)}^2 & = O_P(1),
    \|g^* - \hat g\|_n^2  = O_{P}(n^{-\frac{2\nu}{2\nu + p}}),
    \|\beta^* - \hat \beta\|_2^2  = O_{P}(n^{-\frac{2\nu}{2\nu + p}}).
\end{align*}

\textbf{Case 2:} $\|\hat g\|_{\mathcal{N}_{\Psi}(\Omega)} \leq \|g^*\|_{\mathcal{N}_{\Psi}(\Omega)}$. By \eqref{Pfthm2basicineqX}, \eqref{xbhol} and \eqref{xbhoe}, we have
\begin{align*}
& \|\beta^* - \hat \beta\|_2^2 + \|g^* - \hat g\|_n^2 + \lambda \|\hat g\|_{\mathcal{N}_{\Psi}(\Omega)}^2\nonumber\\
\leq &  O_P(n^{-1/2})\|g^* - \hat g\|_n^{1-\frac{p}{2\nu}} + O_P(n^{-1/2})\|\beta^* - \hat \beta\|_2  + \lambda \|g^*\|_{\mathcal{N}_{\Psi}(\Omega)}^2.
\end{align*}
Using the similar approach as shown in Case 1, it can be shown that
\begin{align*}
    \|\hat g\|_{\mathcal{N}_{\Psi}(\Omega)}^2 & = O_P(1),
    \|g^* - \hat g\|_n^2  = O_{P}(n^{-\frac{2\nu}{2\nu + p}}),
    \|\beta^* - \hat \beta\|_2^2  = O_{P}(n^{-\frac{2\nu}{2\nu + p}}).
\end{align*}
By applying Lemma 5.16 of \cite{geer2000empirical}, we can conclude the asymptotic equivalence of $L_2$ norm and the empirical norm of $\|g^* - \hat g\|_n^2$, i.e.,
\begin{align*}
    \limsup_{n\rightarrow \infty} P\bigg(\sup_{\|g^* - \hat g\|_2^2 \geqslant C_6n^{-\frac{2\nu}{2\nu + p}}, \|\hat g\|_{\mathcal{N}_{\Psi}(\Omega)}^2 = O_P(1) }\bigg|\frac{\|g^* - \hat g\|_n^2}{\|g^* - \hat g\|_{L_2}^2}-1\bigg|\geqslant \eta\bigg) = 0,
\end{align*}
for some constants $C_6$ and $\eta$. Thus, we finish the proof.

\section{Numerical examples}

\subsection{A figure related to Example 1 in Section \ref{secnumconsalg}}

\begin{figure}[h!]
    \centering
    \includegraphics[width=0.5\textwidth]{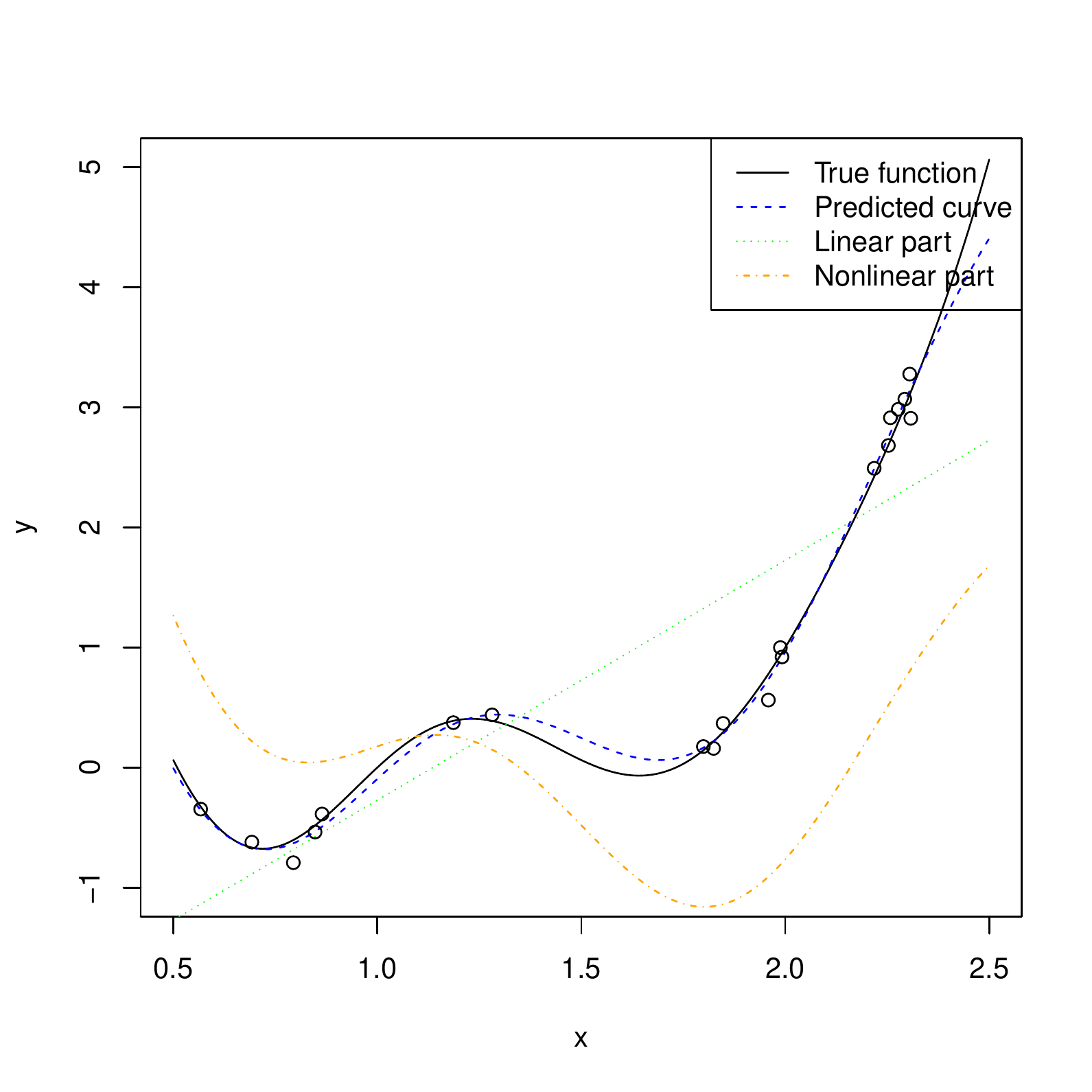}
    \caption{One simulation result of Example 1 in Section \ref{secnumconsalg}. Each dot represents an observation on randomly sampled point.}\label{simple1d}
\end{figure}

Figure \ref{simple1d} shows that the linear part can capture the trend. However, it can be seen from the figure that the difference between the true function and the linear part is still large. Therefore, a nonlinear part is needed to make good predictions. It also indicates that the function in this example is not easy to interpret.

\subsection{Numerical results of Example 2 in Section \ref{secnumconsalg}}
In this section we present the simulation results of Example \ref{eg2num} in Section \ref{secnumconsalg} in the main text. Table \ref{tabnoise01} and Table \ref{tabnoise001} show the simulation results when the variance of noise is 0.1 and 0.01, respectively. We run simulations with iteration numbers $1,2,3,4,5$ for each $n\lambda$, and we find the results are not of much difference. For the briefness, we only present the full simulation results of $n\lambda=1$ to show the similarity, and present the results with $5$ iterations for other values of $n\lambda$.

In Tables \ref{tabnoise01} and \ref{tabnoise001}, we calculate the mean squared prediction error on the training set and the testing set. We also calculate the $L_2$ norm of $\hat f$ and $\hat g$ as in \eqref{num5dobj}, which is approximated by the empirical norm using the first 1000 points of the Halton sequence.
\begin{table}[h]
	\centering
	\begin{tabular}{|c|c|c|c|c|c|}
		\hline
		$n\lambda$ & Iteration number & Training error  & Prediction error & Linear $L_2$ & Nonlinear $L_2$\\
		\hline
		1	& 1 & 0.02951 & 0.01714 &  1.5336 & 0.0034\\
		& 2 & 0.02950 & 0.01712 &  1.5312 & 0.0054 \\
		& 3 & 0.02949 & 0.01711 &  1.5288 & 0.0076\\
		& 4 & 0.02947 & 0.01710 &  1.5265 & 0.0097 \\
		& 5 & 0.02946 & 0.01709 &  1.5242 & 0.0119 \\
		\hline
		0.1	& 5 & 0.02404 & 0.01400 &  1.5264 & 0.02224 \\
		\hline
		0.001 & 5 & 0.0043 & 0.0059 &  1.5285 & 0.1331 \\
		\hline
		$1\times 10^{-9}$	& 5 & $3.860\times 10^{-12}$ & 0.03388 &  1.5324 & 0.2174 \\
		\hline
	\end{tabular}
	\caption{Simulation results when $\epsilon \sim N(0,0.1)$. The third column shows the mean squared prediction error on the training points. The fourth column shows the mean squared prediction error on the testing points. The fifth column and the last column show the approximated $L_2$ norm of $\hat f$ and $\hat g$ as in \eqref{num5dobj}, respectively.}
	\label{tabnoise01}
\end{table}

\begin{table}[h]
	\centering
	\begin{tabular}{|c|c|c|c|c|c|}
		\hline
		$n\lambda$ & Iteration number & Training error  & Prediction error & Linear $L_2$ & Nonlinear $L_2$\\
		\hline
		1	& 	1 & 0.01812 & 0.01759 &  1.5316 & 0.002998\\
		& 	2 & 0.01811 & 0.01757 &  1.5294 & 0.004763 \\
		& 	3 & 0.01810 & 0.01755 &  1.5274 & 0.006664\\
		& 	4 & 0.01809 & 0.01754 &  1.5253 & 0.008581 \\
		& 	5 & 0.01808 & 0.01753 &  1.5234 & 0.010481 \\
		\hline
		0.1	&	5 & 0.01336 & 0.01387 &  1.5203 & 0.017585 \\
		\hline
		0.001 &	5 & 0.00071 & 0.00088 &  1.5287 & 0.120022 \\
		\hline
	\end{tabular}
	\caption{Simulation results when $\epsilon \sim N(0,0.01)$. The third column shows the mean squared prediction error on the training points. The fourth column shows the mean squared prediction error on the testing points. The fifth column and the last column show the approximated $L_2$ norm of $\hat f$ and $\hat g$ as in \eqref{num5dobj}, respectively.}
	\label{tabnoise001}
\end{table}

\bibliographystyle{apalike}
\bibliography{ref}

\end{document}